# Classifying Protoplanetary-disk's Infrared Spectrum and Analysis by c-$C_3H_2$, $C_5H_5$, $C_9H_7$, $C_{12}H_8$, $C_{23}H_{12}$ and $C_{53}H_{18}$ to be Capable Template for Biological Molecules

Norio Ota[1] and Aigen Li[2]

[1]Graduate school of Pure and Applied Sciences, University of Tsukuba, *1-1-1 Tennodai, Tsukuba-City Ibaraki 305-8571, Japan*

[2]Department of Physics and Astronomy, University of Missouri, *Columbia, MO 65211, USA*

Protoplanetary disk around a just born young star contains a lot of cosmic dust. especially polycyclic-aromatic-hydrocarbon (PAH), which would become basic component to create biological organics. This study classified many astronomically observed infrared spectra of protoplanetary disks to three typical spectra. Type-A show well known astronomical bands of 6.2, 7.8, 8.6 and 11.3 micrometer. Whereas Type-B included unknown complex bands. Type-(A+B) was their mixed type. We tried to find specific molecule by Density Functional Theory (DFT) calculation. We found that Type-A could be explained by large PAH molecules of ($C_{23}H_{12}$) and ($C_{53}H_{18}$), which are hexagon-pentagon combined molecular structure. Background molecule of Type-B was smaller ones of (c-$C_3H_2$), ($C_5H_5$), ($C_9H_7$) and ($C_{12}H_8$). Type-(A+B) was reproduced well by mixing those molecules of A and B. Astronomical detailed observation shows that central star of Type-A has larger mass and higher temperature than that of Type-B. This suggests that at very early stage of our solar system, our protoplanetary disk had been made up by Type-B molecules. It was interesting that ($C_5H_5$) and ($C_9H_7$) of Type-B molecules has similar molecular structure with biological nucleic-acid on our earth. Type-B molecules was supposed to become the template for synthesizing biological organics and finally for creating our life.

**Key words**: PAH, protoplanetary disk, infrared spectrum, DFT

## 1. Introduction

Protoplanetary disk was circumstellar dust cloud of just born star younger than 10 million years. Central stars are type of Herbig Ae/Be or T Tauri star. Such protoplanetary disk includes Polycyclic Aromatic Hydrocarbon (PAH), which may had been basic component to create biological organics as like nucleic-acid amino-acid. In this paper, we try to classify many observed infrared spectra of protoplanetary disks by Seok and Li[1]. Also, we like to find specific PAH by Density Functional Theory (DFT) to give identical infrared spectrum to astronomically observed one.

In our recent paper[2], we found specific PAH to explain astronomically well observed mid infrared spectrum by DFT calculation. Astronomical PAH would be floating in interstellar and circumstellar space under ultra-low-density condition of 1~100 molecules/cm$^3$. It is suitable situation for DFT calculation giving solution on such almost isolate molecule[3)4)]. The interstellar gas and dust show common mid-infrared emission at 3.3, 6.2, 7.6, 7.8, 8.6, 11.2, and 12.7μm, which are ubiquitous peaks observed at many astronomical objects[3)-11)]. Current common understanding is that these astronomical spectra come from the vibrational modes of PAH. There are many spectroscopy data[12)-15)] and DFT analysis[16)-21)]. However, despite long-term efforts, until now there is not any identified specific PAH. Our interest was the void-defect induced PAH, which was deformed to a featured structure having few hydrocarbon pentagon rings among hexagon networks. It was surprising that those calculated bands of specific molecules of ($C_{23}H_{12}$) and ($C_{53}H_{18}$) coincided well with astronomically observed bands. Our recent paper[2] would become the first report to indicate the specific PAH in space.

Biological basic molecules may come from interstellar space. To check such idea, we observed and classified infrared spectrum of protoplanetary disks. In this paper, we will classify astronomically observed spectra[1] to three typical spectra of Type-A, Type-B and their mixed Type-(A+B). After that, we like to identify specified PAH by DFT calculation. Especially, small model molecules of $C_6H_6$, $C_{10}H_8$, and $C_{13}H_9$ will give us important information for the background molecules. It will be interesting discussion that void induced and deeply ionized small size PAH may give similar molecular structure with biological molecules and may play the template molecule for creating the nuclear acids and amino acid.

## 2. Classification of observed infrared spectrum

Recently, just born young stars' infrared spectrum attracts many scientists, because it is an analogy of baby age of our solar system and planets. We like to understand how planet system would be created in the Universe. Moreover, it would be key to suppose where and when our biological basic material as like deoxyribonucleic acid and amino acid come from and how synthesized. Astronomically observed infrared spectrum of many protoplanetary disks around the Herbig Ae/Be and T Tauri stars were reported by Acke et al. in 2010[22]. Also in 2017, Seok and Li reported more than 60 observed spectra[1]. Typical examples are noted again in Table 1. Here in this study, we newly classified those spectra to three types, Type-A, Type-B and their mixed type of Type-(A+B) as shown in Fig.1.

In Type-A, we could notice typical astronomical PAH bands, which were marked by blue letter as 6.2μm (a), 7.8μm (b), 8.6μm (c), 11.2μm (d) and 12.7μm (e). Each band wavelength was marked by blue dotted line. On top panel, we can see a typical example of HD97300 (name of central star). In our previous paper[2], those bands could be identified by hydrocarbon pentagon hexagon combined molecules as like $(C_{23}H_{12})^{2+}$ and/or $(C_{53}H_{18})^{1+}$.

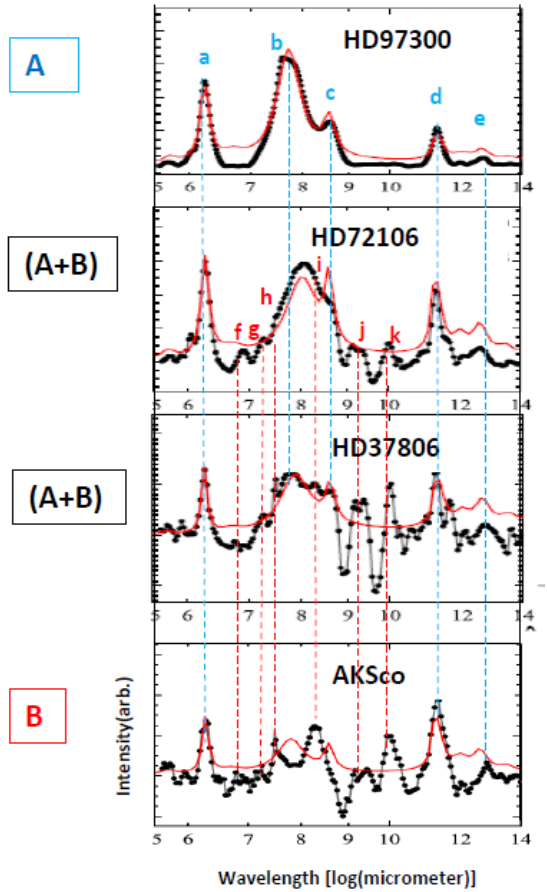

**Fig. 1** Typical observed infrared spectra for Type-A, Type-B and a mixed type of Type-(A+B).

For Type-B, there were complex bands of 6.8μm (red letter "f"), 7.2μm (g), 7.6μm (h), 8.2μm (i), 9.2μm (j) and 10.0μm (k). Bands are marked by red dotted lines. On a bottom column, typical example of AKSco was illustrated. Those bands were not yet identified by any molecule. We like to identify it in this paper. Most protoplanetary disks show mixed type spectrum as Type-(A+B). Examples are HD72106 and HD37806. In Table 1, IR type was marked by A, B and (A+B).

**Table-1** Central stars' characteristics and type of protoplanetary disk's infrared spectrum.

| Object | | IR type | Teff (K) | Ls (Lsun) | Ms (Msun) | Age (Myr) | d (pc) |
|---|---|---|---|---|---|---|---|
| AB | Aur | (A+B) | 9800 | 57.5 | 2.5 | 3.7 | 139 |
| AK | Sco | B | 6500 | 8.9 | 1.66 | 9.3 | 103 |
| BD+40°4124 | | A | 22000 | 5900 | >5.99 | <0.01 | 980 |
| BF | Ori | B | 8750 | 56 | 2.58 | 3.15 | 375 |
| DoAr | 21 | A | 5080 | 7.5 | 2.4 | 0.4 | 120 |
| EC | 82 | B | 4060 | 3.21 | 0.75 | 1.5 | 415 |
| HD | 31648 | (A+B) | 8200 | 15.1 | 1.93 | 7.8 | 137 |
| HD | 34282 | A | 8625 | 13.5 | 1.59 | 6.4 | 191 |
| HD | 34700 | A | 6000 | 20.4 | 1.2 | 10 | 260 |
| HD | 35187 | (A+B) | 8900 | 14.1 | 1.93 | 10.7 | 114 |
| HD | 36112 | B | 7800 | 66 | 2.9 | 2.1 | 279 |
| HD | 36917 | (A+B) | 10000 | 245.5 | 3.98 | 0.72 | 375 |
| HD | 37357 | (A+B) | 9250 | 52.5 | 2.48 | 3.7 | 375 |
| HD | 37411 | (A+B) | 9100 | 34.4 | 1.9 | 9 | 510 |
| HD | 37806 | (A+B) | 11000 | 282 | 3.94 | 0.88 | 375 |
| HD | 38120 | B | 11000 | 41.7 | 2.49 | 5.1 | 375 |
| HD | 58647 | (A+B) | 10500 | 912 | 6 | 1 | 543 |
| HD | 72106 | (A+B) | 11000 | 21.9 | 2.4 | 9 | 289 |
| HD | 85567 | (A+B) | 20900 | 14791 | 12 | ... | 1500 |
| HD | 95881 | (A+B) | 9000 | 7.6 | 1.7 | >3.16 | 118 |
| HD | 97048 | A | 10000 | 44 | 2.53 | 4.61 | 150 |
| HD | 97300 | A | 10700 | 37 | 2.5 | >3 | 188 |
| HD | 98922 | (A+B) | 10500 | 5888 | >4.95 | <0.01 | 1150 |
| HD | 100453 | (A+B) | 7600 | 10 | 1.8 | 15 | 114 |
| HD | 100546 | (A+B) | 10500 | 32 | 2.4 | >10 | 103 |
| HD | 101412 | (A+B) | 8600 | 83 | 3 | 1.2 | 600 |
| HD | 135344B | A | 6750 | 14.5 | 1.9 | 6.6 | 142 |
| HD | 139614 | (A+B) | 7600 | 12.6 | 1.76 | 8.8 | 142 |
| HD | 141569 | A | 9800 | 30.9 | 2.33 | 5.7 | 116 |
| HD | 142527 | (A+B) | 6360 | 23.58 | 2.3 | 2 | 140 |
| HD | 142666 | (A+B) | 7900 | 27.5 | 2.15 | 5 | 145 |
| HD | 144432 | B | 7500 | 19.1 | 1.95 | 6.4 | 145 |
| HD | 145718 | (A+B) | 8100 | 19.5 | 1.93 | 7.4 | 145 |
| HD | 163296 | B | 9200 | 33 | 2.23 | 5.1 | 119 |
| HD | 169142 | A | 8250 | 8.55 | 1.69 | 6 | 145 |
| HD | 179218 | (A+B) | 9640 | 182 | 3.66 | 1.08 | 254 |
| HD | 244604 | (A+B) | 8200 | 55 | 2.66 | 2.79 | 375 |
| HD | 250550 | (A+B) | 11000 | 138 | 3.1 | 1.42 | 700 |
| HD | 281789 | (A+B) | 9520 10c | ... | ... | 350 | |
| IC348 | LRL110 | (A+B) | 3778 | 0.22 | 0.78 | ... | 250 |
| IRAS | 03260+3111 | A | 13000~180d | | ... | ... | 290 |
| IRAS | 06084-0611 | A | 18600e 10c | | ... | ... | 1050 |
| J182858.1+001724 | | (A+B) | 5830 | 4.76 | 1.4 | ... | 260 |
| J182907.0+003838 | | (A+B) | 4060 | 0.6 | 1.04 | ... | 260 |
| LkHα | 224 | (A+B) | 7850 | 115 | 6.05 | 0.312 | 980 |
| LkHα | 330 | B | 5800 | 11 | 2.5 | 3 | 250 |
| MWC | 297 | (A+B) | 24000 | 10230 | 9 | 1 | 250 |
| MWC | 865 | A | 14000 | 8710 | 9 | 1 | 400 |
| MWC | 1080 | A | 30000 | 180000 | 10 | 1 | 2200 |
| Oph | IRS48 | A | 10000 | 14.3 | 2.25 | 15 | 120 |
| PDS | 144N | (A+B) | 8750 10c | | ... | ... | 1000 |
| RR | Tau | A | 8460 | 2 | 3.57 | 1.34 | 160 |
| RXJ1615.3-3255 | | (A+B) | 4590 | 0.85 | 1.28 | ... | 120 |
| SR | 21N | A | 5830 | 6.5 | 1.7 | 1 | 125 |
| SU | Aur | (A+B) | 5860 | 11 | 1.88 | 6.3 | 146 |
| TY | CrA | A | 12000 | 98 | 3.16 | 3 | 140 |
| UX | Tau | (A+B) | 5520 | 3.5 | 1.5 | 3 | 140 |
| V590 | Mon | (A+B) | 13000 | 295 | 4.71 | 2.8 | 800 |
| V892 | Tau | A | 8000 | 21 | 2.06 | 6.11 | 140 |
| VV | Ser | (A+B) | 14000 | 324 | 4 | 0.64 | 260 |
| WL | 16 | A | 9000 | 250 | 4 | 1 | 125 |
| Wray | 15-1484 | A | 30000 | 977 | ... | ... | 750 |
| WW | Vul | B | 9000 | 170 | 3.7 | 0.9 | 700 |

It is interesting to map those IR types in relation of central star's mass (solar unit) and effective temperature Teff (K). Type-A (blue dot) takes a wide range of Mass and Teff as surrounded by blue circle, whereas Type-B has lighter and lower temperature central star. Our early solar system would be Type-B as an object of 1 solar mass and effective temperature of 4000~5000K.

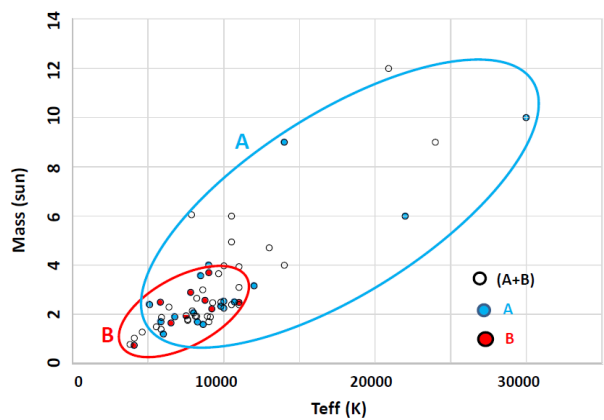

**Fig. 2** Classifying IR types in central star's mass (solar unit) vs. effective temperature.

### 3. Candidates for Background Molecules

#### 3.1 Model molecules

This paper focuses to find background molecules for explaining spectra of Type-B and Type-(A+B). It is common understand that, as illustrated in Fig. 3, just born young star emit high energy electron and proton, also emit high energy photon, which may attack hydrocarbon molecules hidden in a cloud of protoplanetary disk.

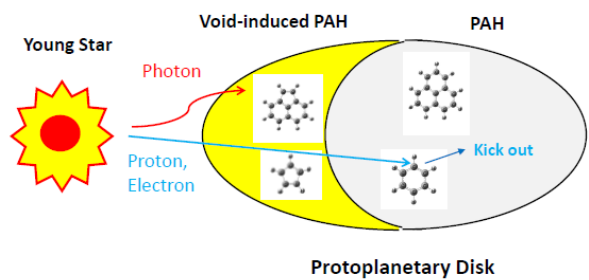

**Fig. 3** Image of young central star, emitting electron, proton and photon to attack on PAH molecules in the cloud of protoplanetary disk.

Model molecules hidden in the protoplanetary disk are illustrated in Fig. 4. Starting mother molecules are typical PAH of ($C_6H_6$), ($C_{10}H_8$), ($C_{13}H_9$), ($C_{24}H_{12}$), and ($C_{54}H_{18}$) having hydrocarbon hexagon rings. In this paper, we apply one assumption of single void-defect on initial molecule. In interstellar and circumstellar space, high speed stellar wind, mainly proton and electron, may attack on PAH. As illustrated on top of Fig. 3, high speed particle attacks mother molecule ($C_6H_6$) and kick out one carbon atom. Initial void will be created and immediately make a void induced molecule as like ($C_5H_5$). For every size initial molecule, DFT calculation suggested serious structure change. For example, in case of ($C_{23}H_{12}$) two carbon pentagon rings are created among hexagon-ring network. For those hydrocarbon pentagon hexagon combined molecules, we also supposed photoionization by central stars, which will make deep photoionization for hydrocarbon pentagon hexagon combined molecules.

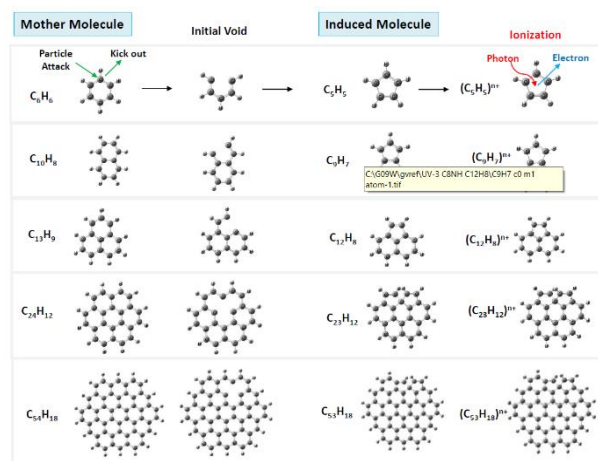

**Fig. 4** Creation of model molecules. Mother molecule will be attacked by high energy particle from the central star making initial void. Void induced molecule will be illuminated by central star and makes photoionized cations.

#### 3.2 Calculation methods

In calculation, we used DFT[22),23)] with the unrestricted B3LYP functional[24)]. We utilized the Gaussian09 software package[25)] employing an atomic orbital 6-31G basis set[26)]. Unrestricted DFT calculation was done to have the spin dependent atomic structure. The required convergence of the root-mean-square density matrix was $10^{-8}$. Based on such optimized molecular configuration, fundamental vibrational modes were calculated, such as C-H and C-C stretching modes, C-H bending modes and so on, using the same Gaussian09 software package. This calculation also gives harmonic vibrational frequency and intensity in infrared region. The standard scaling is applied to the frequencies by employing a scale factor of 0.965 for PAH from the laboratory experimental value on coronene ($C_{24}H_{12}$)[27)]. Correction due to anharmonicity was not applied to avoid uncertain fitting parameters. To each spectral line, we

## 4. Photoionization from ($C_5H_5$) to ($c$-$C_3H_2$) and chain-$C_3$

Astrochemical evolution step of small PAH started from benzene ($C_6H_6$) was illustrated in Fig.5. One astronomical hypothesis is high speed particle attack on mother molecule ($C_6H_6$) to create a void defect. Void induced open configuration would be suddenly transformed to a cyclic hydrocarbon pentagon ($C_5H_5$). Next hypothesis is photo-irradiation from the central star introducing deep photoionization on ($C_5H_5$). Neutral ($C_5H_5$) will be excited to mono-cation ($C_5H_5$)$^{1+}$ by 8.5eV, to di-cation ($C_5H_5$)$^{2+}$ by 23.6eV, also tri-cation ($C_5H_5$)$^{3+}$ by 46.1eV. Calculation shows that from n=0 to 3 of ($C_5H_5$)$^{n+}$, molecule keeps pentagon configuration. At ionization step n=4, there occurs dehydrogenation to cyclic pure carbon ($C_5$) and five hydrogen atoms. At n=6 shown in a green frame, there happens surprising creation of cyclic ($c$-$C_3H_2$) marked by red ellipse, also separation to two carbon atoms and three hydrogen atoms. ($c$-$C_3H_2$) is the smallest PAH.

Calculated infrared spectrum of ($C_5H_5$)$^{n+}$ (n=0, 1, and 2) were illustrated in Fig. 6. Stable spin state was $S_z=1/2$ for cases of neutral ($C_5H_5$)$^0$ and di-cation ($C_5H_5$)$^{2+}$, of which spin cloud was shown by red for up-spin and by blue for down-spin at density surface of $10e/nm^3$. As shown in Fig. 7, deeper photoionization on ($c$-$C_3H_2$) creates chain-$C_3H_2$ and finally decompose to pure carbon $C_3$ and two hydrogen atoms.

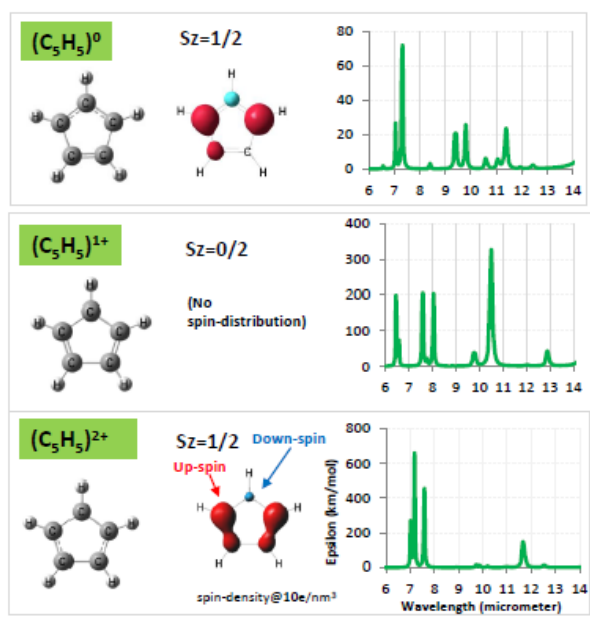

Fig. 6 Atomic configuration and infrared spectrum of ionized ($C_5H_5$)$^{n+}$ for n=0,1 and 2. Spin density was illustrated by red for up-spin and blue for down spin.

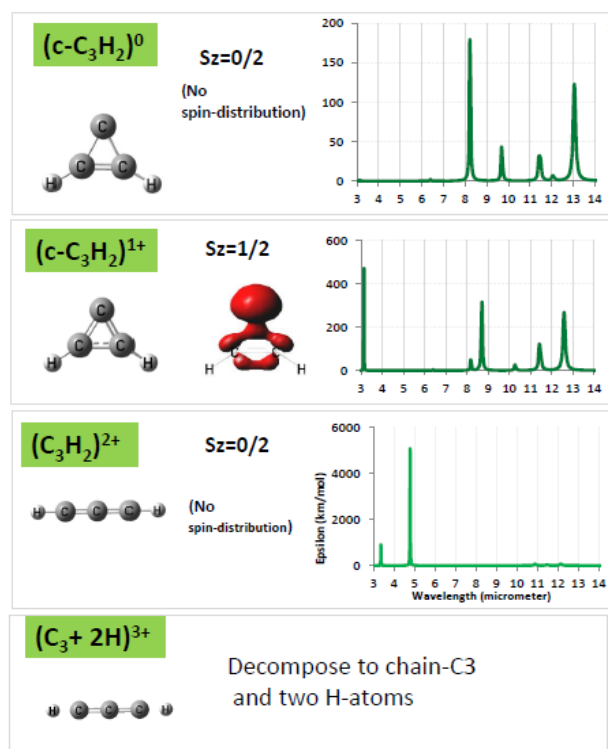

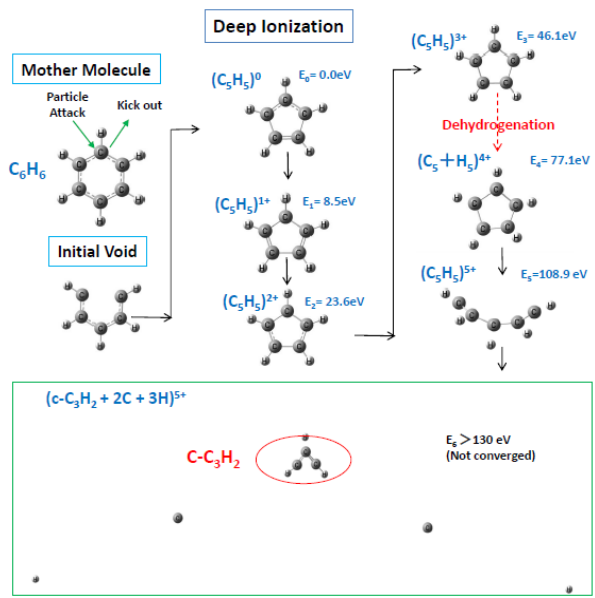

Fig. 5 Deep photoionization of void induced molecule ($C_5H_5$) to create the smallest PAH ($c$-$C_3H_2$)

Fig. 7 Deeper ionization from $c$-$C_3H_2$ to chain-$C_3$

## 5. Large PAHs: (C₉H₇), (C₁₂H₈), (C₂₃H₁₂) and (C₅₃H₁₈)

To find PAH explaining Type=B infrared spectrum, we expand DFT calculation to various size molecules.

(1) ($C_9H_7$)

Starting from mother molecule of ($C_{10}H_8$), void induced molecule ($C_9H_7$) was studied. As illustrated in Fig. 8, deeply ionized molecule show dehydrogenation at a step of ($C_9H_7$)$^{6+}$, and finally decomposed to several chain-hydrocarbons and atoms. Calculated atomic configuration and infrared spectrum of ionized ($C_9H_7$)$^{n+}$ for n=0, 1 and 2 were shown in Fig. 9.

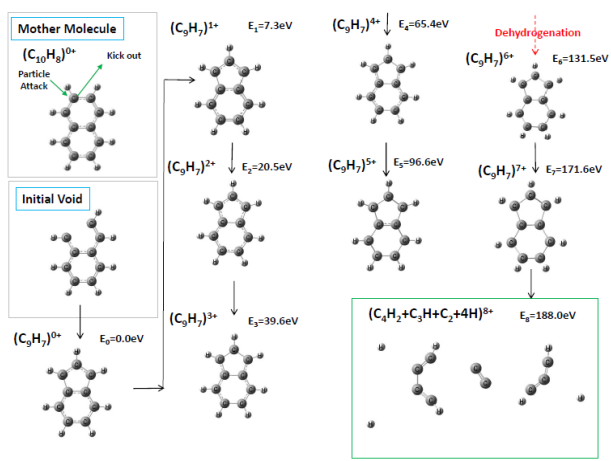

**Fig. 8** Deep ionization step of ($C_9H_7$).

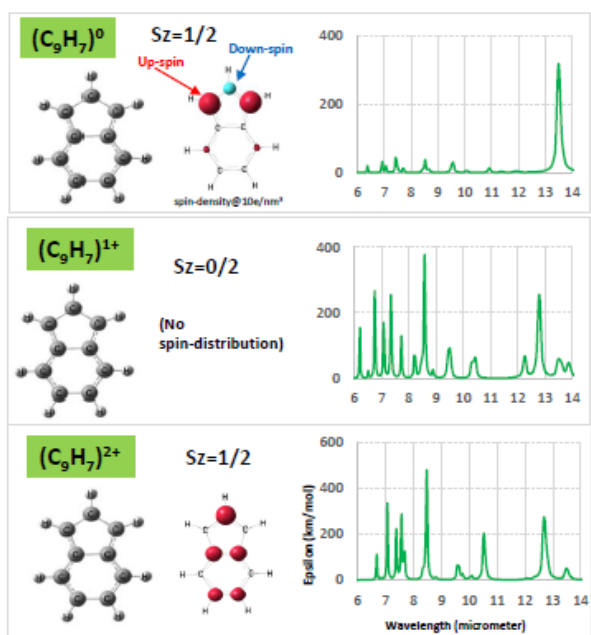

**Fig. 9** Atomic configuration and spin distribution infrared spectrum of ionized ($C_9H_7$)$^{n+}$ for n=0, 1 and 2.

(2) ($C_{12}H_8$)

Void induced molecule ($C_{12}H_8$) was created from mother molecule ($C_{13}H_9$). As illustrated in Fig. 10, deeply ionized molecule shows de-hydrogenation and decomposition to complex chain-hydrocarbon. Fig. 11 show calculated atomic configuration, spin distribution and infrared spectra.

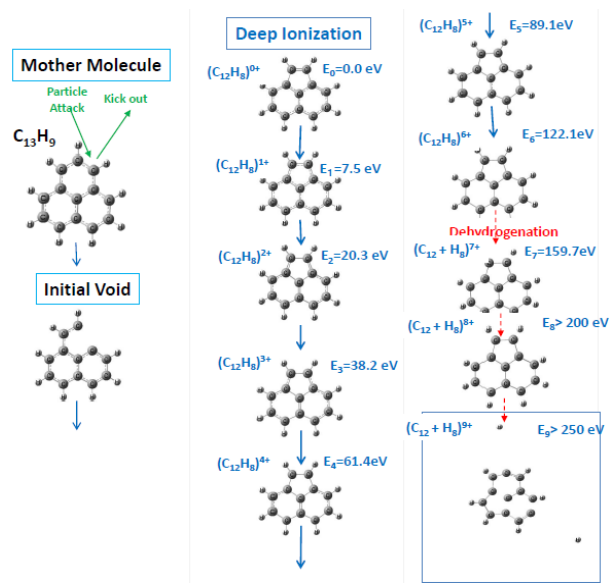

**Fig. 10** Ionization step of ($C_{12}H_8$).

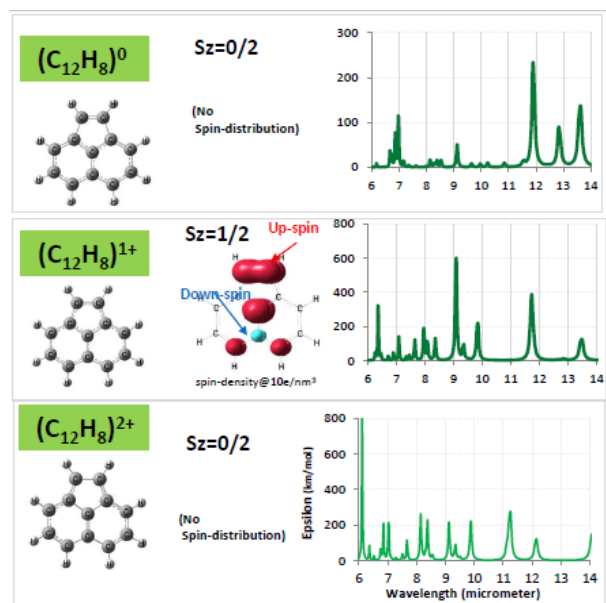

**Fig. 11** Calculated atomic configuration, spin density and infrared spectrum of ionized ($C_{12}H_8$)$^{n+}$.

(3) $(C_{23}H_{12})$

Void induced molecule of $(C_{23}H_{12})$ was created from mother molecule $(C_{24}H_8)$, which was well studied and reported in a previous paper[2]. Fig. 12 show calculated atomic configuration and infrared spectra of $(C_{23}H_{12})^{n+}$ from n=0 to 2.

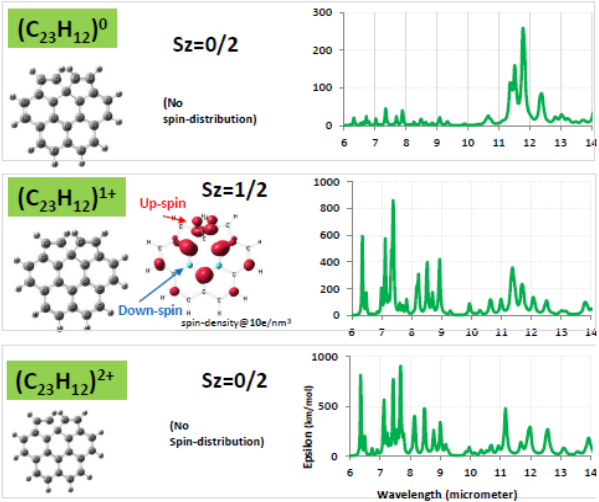

**Fig. 12** Calculated atomic configuration, spin density and infrared spectrum of ionized $(C_{23}H_{12})^{n+}$.

(4) $(C_{53}H_{18})$

Void induced molecule of $(C_{53}H_{18})$ was well studied in a previous paper[2]. Fig. 13 show calculated atomic configuration, spin density and infrared spectrum.

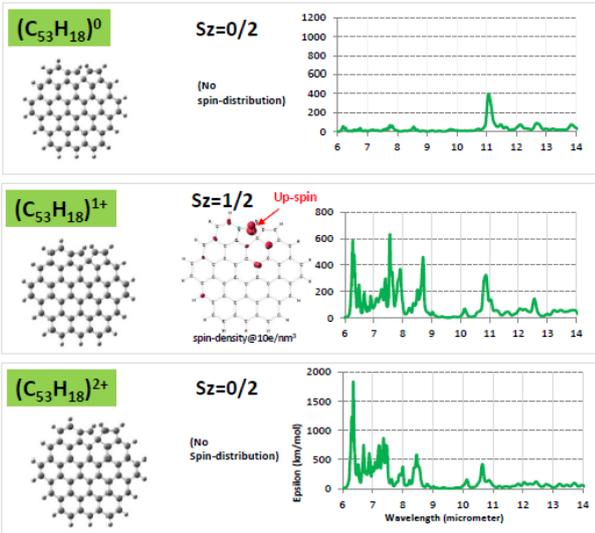

**Fig. 13** Calculated atomic configuration, spin density and infrared spectrum of ionized $(C_{53}H_{18})^{n+}$.

Detailed infrared bands were analyzed as shown in Fig. 14. For example, atomic vibrational mode of 6.2μm band was carbon-to-carbon stretching mode and for 7.6μm band carbon-to-hydrogen bending mode.

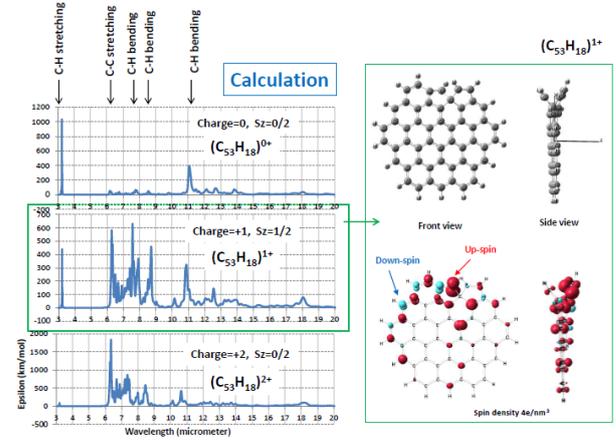

**Fig. 14** Detailed atomic configuration and spin density map of mono-cation $(H_{53}H_{18})^{1+}$. For every band, vibrational modes were analyzed as C-C stretching or C-H bending as marked by arrow.

## 6. Molecules for reproducing Type-A spectrum.

We tried to find suitable molecules for reproducing Type-A astronomically observed spectrum. Fig. 15 show comparison with the Type-A observed spectrum of HD97300 (central star) and the calculated spectra of various size molecules. For easy comparison, red dotted arrow shows no good coincident wavelength between Type-A observed band and DFT calculated one. In cases of $(c\text{-}C_3H_2)^0$ and $(C_5H_5)^0$, we cannot find any coincidence between observation and calculation. Whereas in case of medium size molecules of $(C_9H_7)^{1+}$ and $(C_{12}H_8)^{2+}$, we noticed many non- coincident bands as checked by red arrows, but find some coincident bands as like at 6.2, 7.6 and 8.6μm. It should be noted that larger molecules of $(C_{23}H_{12})^{2+}$ show good coincidence between calculated bands and observed Type-A bands. Especially, the largest molecule of $(C_{53}H_{18})^{1+}$ show very good coincidence with Type-A. Fig. 16 shows detailed charge state variation of calculated spectrum of $(C_{53}H_{18})^{n+}$ for cases of N=0,1 and 2. Observed spectrum may be some weighting sum of those charge state.

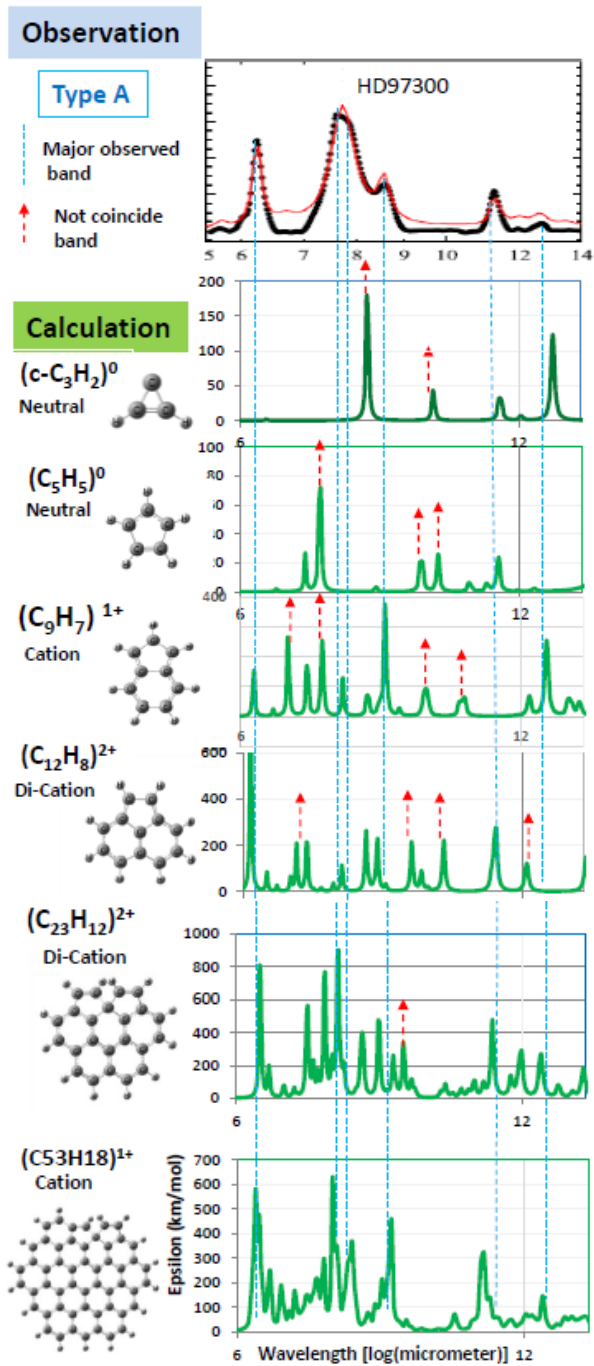

**Fig. 15** Comparison with calculated infrared bands with Type-A observed bands. Red arrow shows no good coincident band.

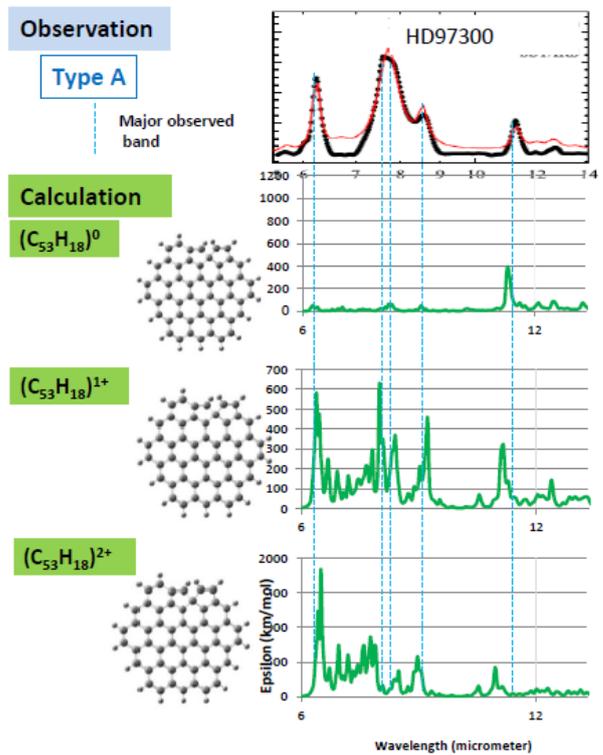

**Fig. 16** Observed infrared spectrum of Type A of star HD97300 on a top panel, which were compared with calculated infrared spectrum of $(C_{53}H_{18})^{n+}$ (n=0, 1 and 2).

## 7. Molecules reproducing Type-B spectrum.

We tried to find suitable molecules for reproducing Type-B astronomically observed spectrum. Fig. 17 show comparison with the Type-B spectrum of AKSco (central star) and the calculated spectra of various size molecules. In cases of $(c\text{-}C3H2)^0$ and $(C5H5)^0$, we could find that calculated four or five bands were well reproduce well Type-B bands. It should be noted that in case of medium size molecules of $(C_9H_7)^{1+}$ and $(C_{12}H_8)^{2+}$, calculated bands show complete reproduction of observed 12 bands of Type-B. Whereas for larger molecules of $(C_{23}H_{12})^{2+}$ and $(C_{53}H_{18})^{1+}$, we noticed many non-coincident bands as marked by red dotted arrows. It is obvious that background molecules of Type-B were a sum of smaller and medium size molecules. Fig. 18 shows detailed charge state variation of calculated spectrum of $(C_{12}H_8)$. Observed spectrum may be some weighting sum of those charge state.

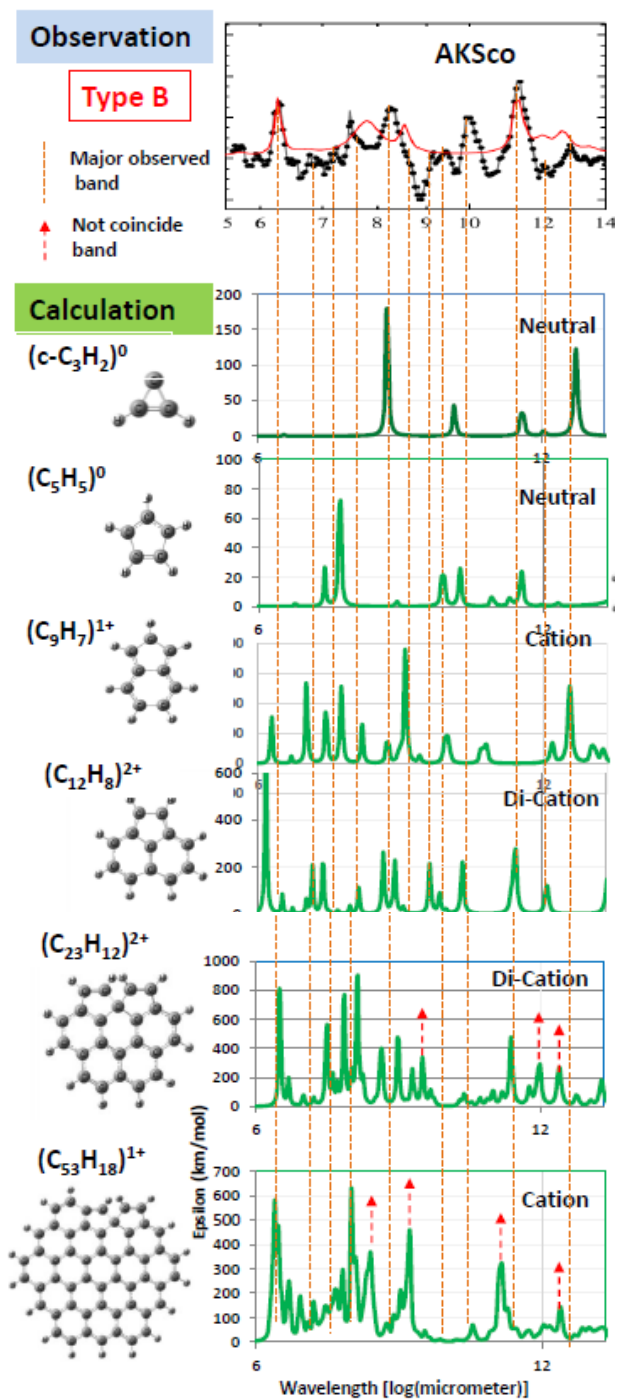

**Fig. 17** Comparison with calculated infrared bands with Type-B observed bands. Red arrow shows no good coincident band.

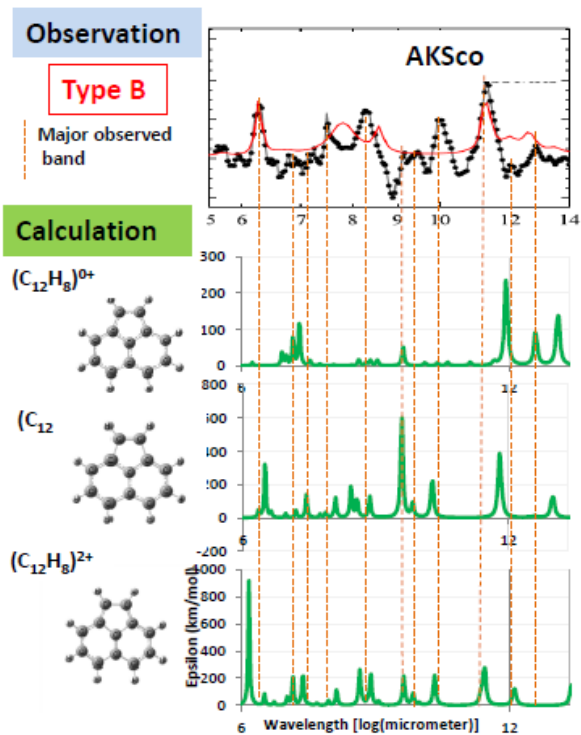

**Fig. 18** Charge state variation of calculated infrared spectrum of $(C_{12}H_8)$.

## 8. Reproducing Type-(A+B) spectrum.

As noted in section 2, astronomically observed Type-(A+B) would be a sum of observed Type-A and Type-B. We tried to reproduce Type-(A+B) by using two typical molecules, that is, $(C_{23}H_{12})^{2+}$ for Type-A and $(C_{12}H_8)^{2+}$ for Type-B. On a top panel of Fig. 19, observed spectrum of HD72106 could be well reproduced by a sum of 60% of A and 40% of B. Also, HD142527 was well reproduced by a sum of 40% of A and 60% of B. Finally, HD37806 was reproduced by a sum of 20% of A and 80% of B.

## 9. Capable template molecules for creating primitive biological components.

Small molecules of $(C_5H_5)$ and $(C_{12}H_8)$ would be background molecules for Type-B spectrum. Our solar system had been Type-B at the baby age. On the very early stage on the earth, biological basic molecules would be created by some chemical revolution mechanism. Here, we could suppose one interesting hypothesis that cosmic PAH dust may become the template for creating primitive biological molecules as like Cytosine:$(C_4H_5N_3O)$ and Guanine:$(C_5H_5N_5O)$. As illustrated in Fig. 19, we could notice resemblance of molecular configuration between cosmic dust PAH and biological nucleic acid. In the water of pond/sea, we could suppose chemical revolution step by introducing some contents of $NH_3OH$ under high temperature environment of hot springs and/or hot spot. Of course, we need detailed experiment on earth and study to find astronomical evidence.

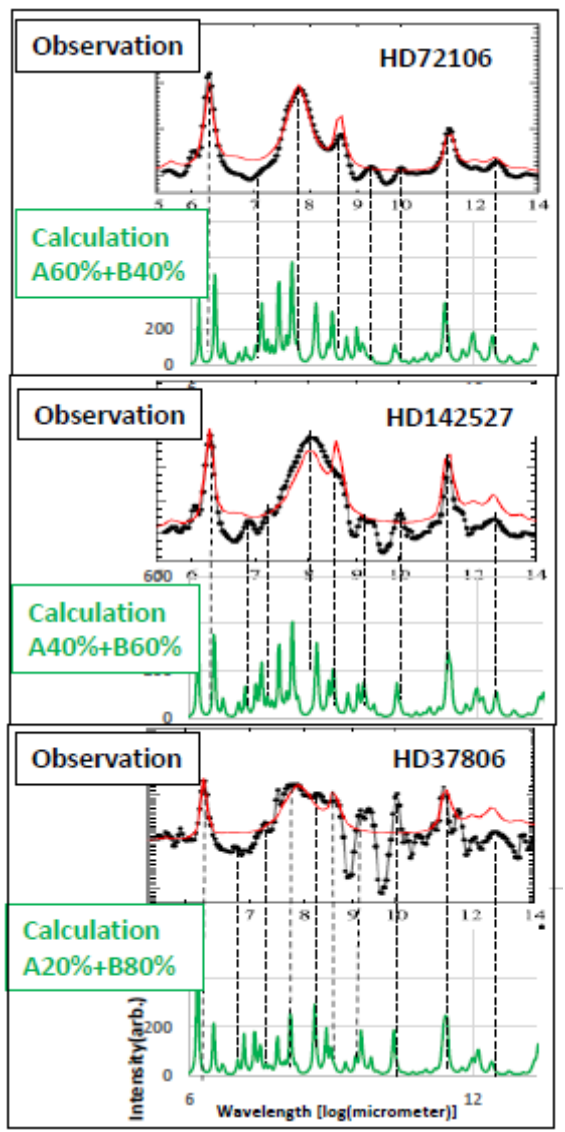

**Fig. 19** Reproduction of Type-(A+B) observed spectrum by a sum of two typical molecules of $(C_{23}H_{12})^{2+}$ for A, and $(C_{12}H_8)^{2+}$ for B.

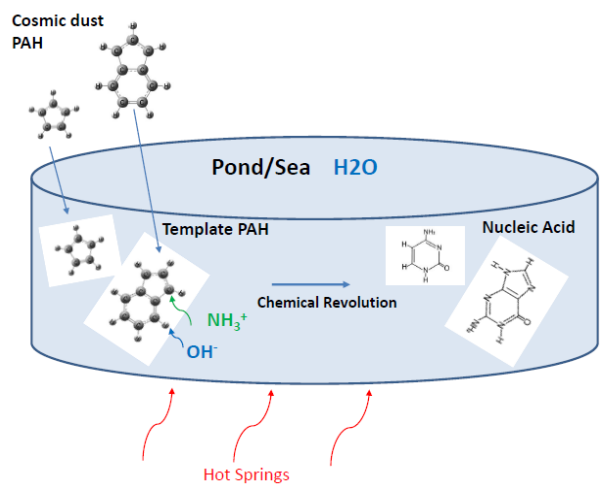

**Fig. 20** Image of creating biological components on very early planet. Cosmic PAH dust would become the template molecule for creating basic biological molecules.

## 10. Conclusion

It is important to find specific molecule included in just born star's protoplanetary disks, especially specific PAH, which would become primitive component to create biological organics.

(1) Many astronomically observed infrared spectra of protoplanetary disks by Seok & Li[1] were classified to three typical spectra. Type-A show well known bands of 6.2, 7.8, 8.6 and 11.3 micrometer. Whereas Type-B was unknown complex bands and Type-(A+B) their mixed bands.

(2) We tried to find specific molecule by Density Functional Theory (DFT). Model molecules were various size mother PAHs starting from $C_5H_5$, $C_{10}H_8$, $C_{13}H_9$, $C_{23}H_{12}$ and $C_{54}H_{18}$. By the central star, those mother molecules would be attacked by high energy particles and photon, to be deeply ionized pentagon hexagon combined PAH.

(3) We found that Type-A could be explained by large molecules of $(C_{23}H_{12})$ and $(C_{53}H_{18})$.

(4) Background molecule of Type-B was smaller ones of $(c\text{-}C_3H_2)$, $(C_5H_5)$, $(C_9H_7)$ and $(C_{12}H_8)$. Type-(A+B) was well reproduced by mixing those molecules.

(5) Central star's mass vs. effective temperature was mapped. Star of Type-A show larger mass and higher temperature than that of Type-B. At very early stage of our solar system (Teff~5000K, 1 solar mass) may be Type-B.

(6) It was interesting that $(C_5H_5)$ and $(C_9H_7)$ has similar molecular structure with biological nuclear acid on our earth. Background molecules of Type-B would become the template molecule for synthesizing biological organics and finally creating our life.

## Acknowledgement

Aigen Li is supported in part by NSF AST-1311804 and NASA NNX14AF68G.